\documentclass[aps,prl,reprint,superscriptaddress,floatfix,amssymb,longbibliography]{revtex4-1}

\usepackage{graphicx}
\usepackage{dcolumn}
\usepackage{amsmath}
\usepackage{amssymb}
\usepackage[usenames, dvipsnames]{color}

\usepackage[pdftex]{hyperref}
\hypersetup{colorlinks=true,linkcolor=blue,citecolor=blue,urlcolor=blue}

\begin{document}

\title{${}^{239}$Pu nuclear magnetic resonance in the candidate topological insulator PuB$_4$}

\author{A. P. Dioguardi}
\email{adioguardi@gmail.com}
\affiliation{Los Alamos National Laboratory, Los Alamos, New Mexico 87545, USA}
\affiliation{IFW Dresden, Institute for Solid State Research, P.O. Box 270116, D-01171 Dresden, Germany}

\author{H. Yasuoka}
\affiliation{Los Alamos National Laboratory, Los Alamos, New Mexico 87545, USA}
\affiliation{Max Planck Institute for Chemical Physics of Solids, 01187 Dresden, Germany}

\author{S. M. Thomas}
\affiliation{Los Alamos National Laboratory, Los Alamos, New Mexico 87545, USA}

\author{H. Sakai}
\affiliation{Advanced Science Research Center, Japan Atomic Energy Agency, Tokai, Naka, Ibaraki 319-1195, Japan}
\affiliation{Los Alamos National Laboratory, Los Alamos, New Mexico 87545, USA}

\author{S. K. Cary}
\author{S. A. Kozimor}
\affiliation{Los Alamos National Laboratory, Los Alamos, New Mexico 87545, USA}

\author{T. E. Albrecht-Schmitt}
\affiliation{Department of Chemistry and Biochemistry, Florida State University, 95 Chieftan Way, Tallahassee, Florida 32306}

\author{H. C. Choi}
\author{J.-X. Zhu}
\author{J. D. Thompson}
\author{E. D. Bauer}
\author{F. Ronning}
\affiliation{Los Alamos National Laboratory, Los Alamos, New Mexico 87545, USA}

\date{\today}

\begin{abstract}
We present a detailed nuclear magnetic resonance (NMR) study of ${}^{239}$Pu in bulk and powdered single-crystal plutonium tetraboride (PuB$_4$), which has recently been investigated as a potential correlated topological insulator. This study constitutes the second-ever observation of the ${}^{239}$Pu NMR signal, and provides unique on-site sensitivity to the rich $f$-electron physics and insight into the bulk gap-like behavior in PuB$_4$. The ${}^{239}$Pu NMR spectra are consistent with axial symmetry of the shift tensor showing for the first time that ${}^{239}$Pu NMR can be observed in an anisotropic environment and up to room temperature. The temperature dependence of the ${}^{239}$Pu shift, combined with a relatively long spin-lattice relaxation time ($T_1$), indicate that PuB$_4$ adopts a non-magnetic state with gap-like behavior consistent with our density functional theory (DFT) calculations. The temperature dependencies of the NMR Knight shift and $T_1^{-1}$---microscopic quantities sensitive only to bulk states---imply bulk gap-like behavior confirming that PuB$_4$ is a good candidate topological insulator. The large contrast between the ${}^{239}$Pu orbital shifts in the ionic insulator PuO$_2$ ($\sim +24.7$~\%) and PuB$_4$ ($\sim -0.5$~\%) provides a new tool to investigate the nature of chemical bonding in plutonium materials. 
\end{abstract}

\maketitle

Topological insulators have received much attention recently due to the experimental verification of the theoretical prediction of topologically nontrivial symmetry-protected surface states~\cite{Hasan:2010ti,Qi:2011ti}. Kondo insulators are $f$-electron systems with strong correlations in which hybridization of the $f$-electrons with conduction electrons forms a gap at the Fermi level~\cite{Coleman:2007hf}. Strong spin-orbit coupling can result in a topological Kondo insulator in which band inversion drives the emergence of nontrivial topologically protected gapless surface states~\cite{Dzero:2010tk, Dzero:2015tk}. Samarium hexaboride (SmB$_6$) is the primary candidate example of a topological Kondo insulator~\cite{Takimoto:2011sm, Kim:2013sm, Neupane:2013sm, jiang2013observation}. As compared with rare-earth $4f$-electron systems, the actinide $5f$-electron systems have more spatially-extended $f$-electron wave functions, which generally results in an enhancement of the energy scales involved~\cite{Moore:2009am, clark2000chemical, Deng:2013bb}. Plutonium (Pu) materials display particularly complex physical properties due to the $5f$-electrons lying on the brink between bonding and non-bonding configurations~\cite{Shim:2007er, Janoschek:2015fg}. For example, elemental Pu forms in six allotropes at ambient pressure that vary in density by up to 25\%~\cite{Clark:2006fy, Lashley:2005gs}. Pu compounds display a wide variety of electronic ground states including heavy-fermion behavior, magnetism, superconductivity~\cite{Bauer:2015jo}, and most recently the prediction of topologically non-trivial states~\cite{Deng:2013bb,Zhang:2012ac}.

\begin{figure}[t!] 
	\includegraphics[trim=0cm 0cm 19cm 0cm, clip=true, width=0.9\linewidth]{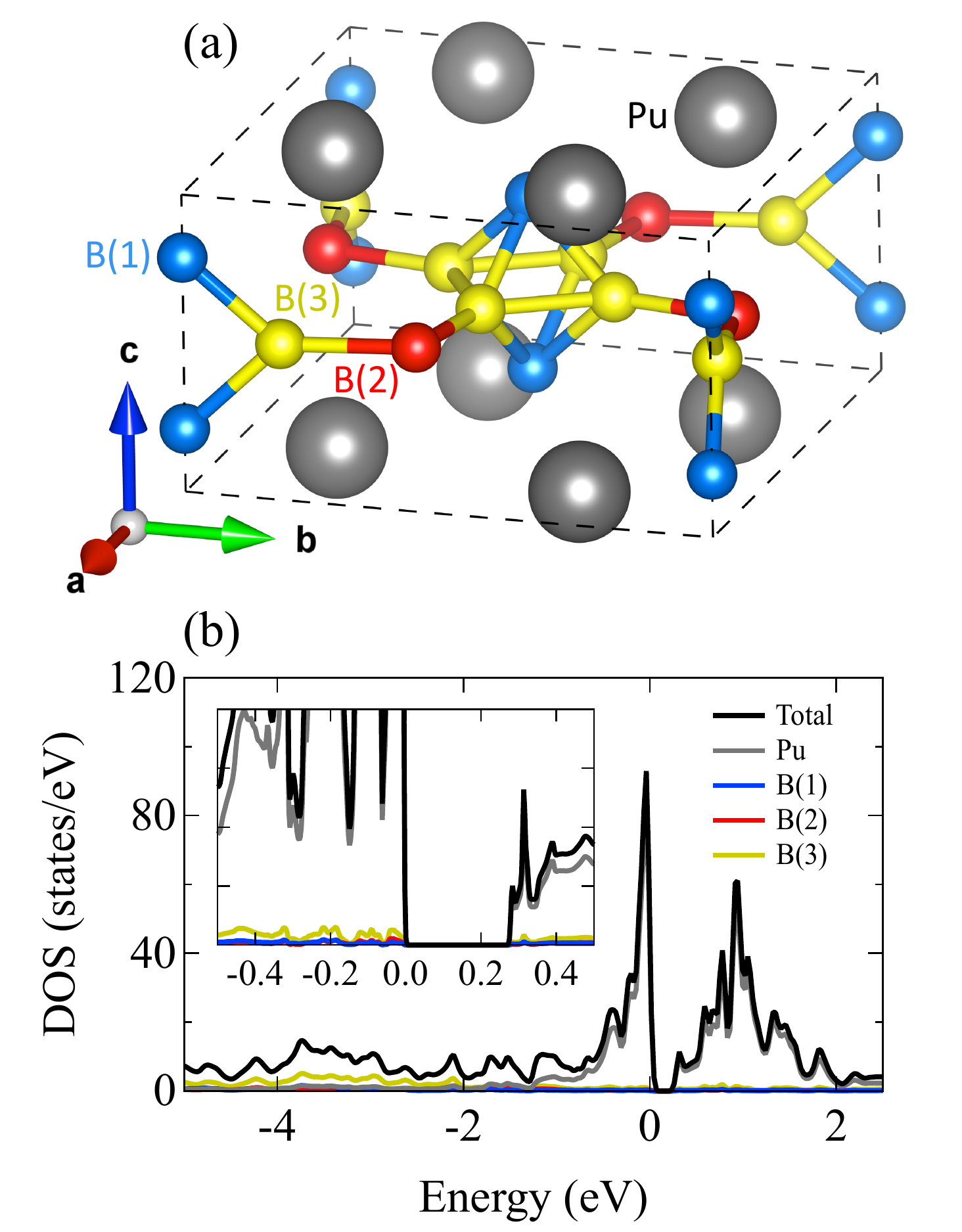}
	{\caption{\label{fig:PuB4_unit_cell_DOS}(a) Unit cell of PuB$_4$ illustrating the single plutonium site and three inequivalent boron sites~\cite{Momma:2011dd}. (b) Density of states (DOS) and partial DOS as a function of energy calculated within density functional theory including spin-orbit coupling. The inset shows an expanded region near the Fermi energy where there is an energy gap of $\Delta \sim 254$ meV.}}
\end{figure}

Very recently, plutonium tetraboride (PuB$_4$) has been theoretically predicted to be a strong topological insulator in which electronic correlations play an important role~\cite{Choi:2018up}. The density functional theory (DFT) calculations predict a band gap $\Delta \sim 254$ meV and dynamical mean-field theory (DMFT) calculations find that electronic correlations significantly reduce the magnitude of the predicted energy gap. Experimental measurements from the same work find an increase of the resistivity with decreasing temperature and saturation at low temperature reminiscent of the behavior of SmB$_6$~\cite{Cooley:1995sm}. Fits to the temperature dependent resistivity yield an energy gap $\Delta = 35$ meV, which is taken as evidence for correlation-induced suppression of the expected gap value. PuB$_4$ forms in the tetragonal ThB$_4$-type crystal structure with space group \textit{P4/mbm} (\# 127) as shown in Fig.~\ref{fig:PuB4_unit_cell_DOS}(a) and was first reported nearly 60 years ago~\cite{McDonald:1960kj,Eick:1965gk,Rogl:1997cx}. Magnetic measurements of PuB$_4$ indicated that the Pu magnetic moment is very small, on the order of $7.2 \times 10^{-4}$ emu/mol and shows little temperature dependence~\cite{Smith:gj}. This small magnetic susceptibility and insulating-like electrical transport make PuB$_4$ an ideal material in which to search for ${}^{239}$Pu nuclear magnetic resonance (NMR).

\begin{figure*}[t!] 
	\includegraphics[trim=0cm 0cm 0cm 0cm, clip=true, width=\linewidth]{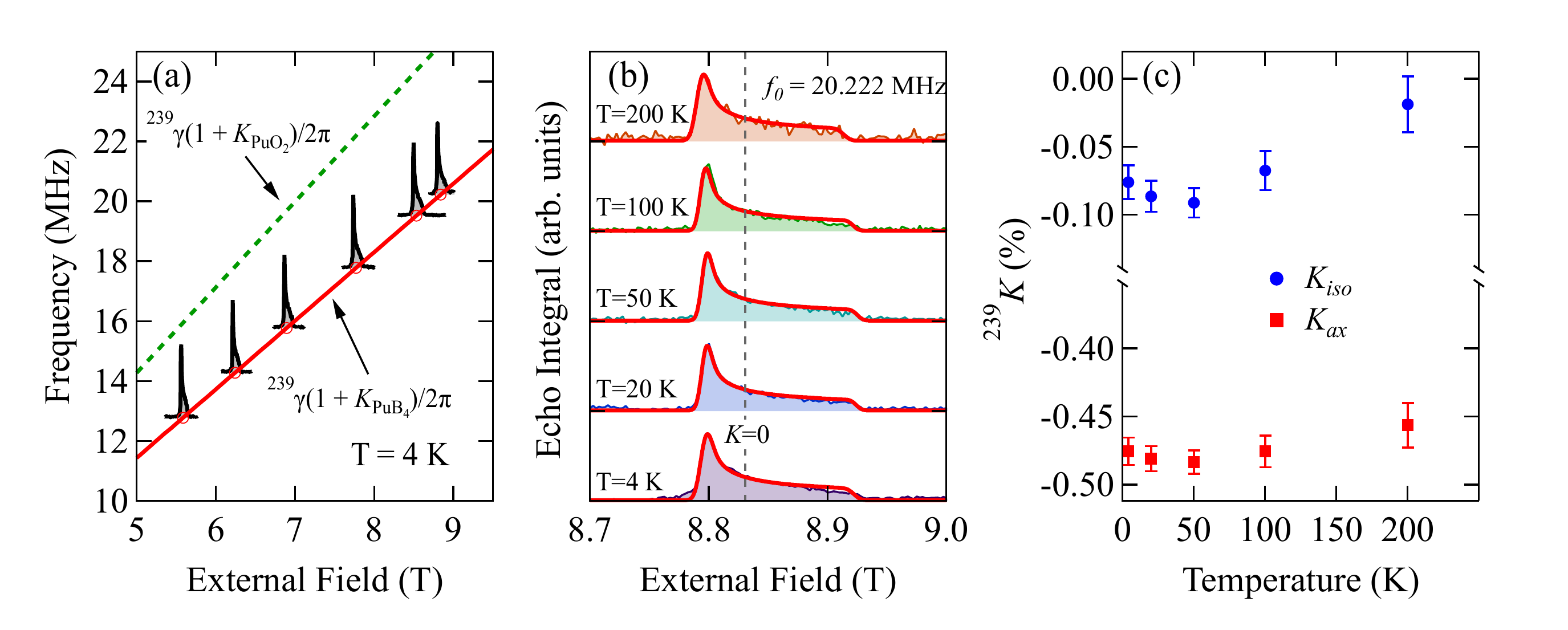}
	\caption{\label{fig:Pu239_spectra_layout}(a) ${}^{239}$Pu nuclear magnetic resonance (NMR) field-swept spectra of powdered single crystals of PuB$_4$ at several frequencies at $T=4$ K. Spectra are normalized to the maximum value and offset vertically so as to correspond to the observed frequency $f_0$ on the left axis. Red circles and line indicate the resonant condition of ${}^{239}$Pu in PuB$_4$ ${}^{239}\gamma(1 + K_{\mathrm{PuB}_4})/2\pi = 2.288 \pm 0.001$ MHz/T (at $T=4$ K) and green dashed line shows ${}^{239}\gamma(1 + K_{\mathrm{PuO}_2})/2\pi = 2.856 \pm 0.001$ MHz/T as determine previously~\cite{Yasuoka:2012fc}. (b) ${}^{239}$Pu NMR field-swept spectra at $f_0 = 20.222$ MHz offset vertically for several temperatures. Solid red curves are best fits as described in the text. Vertical dashed line indicates zero shift $K=0$ using  ${}^{239}\gamma/2\pi = 2.29$ MHz/T. (c) ${}^{239}K_{iso}$ and ${}^{239}K_{ax}$ vs temperature extracted from fits in (b).}
 \end{figure*}

NMR is a powerful tool for the investigation of the physics and chemistry of condensed matter in general~\cite{slichter:1990pr, Curro2016NMRKondo, Kinross:2014ea, ashbrook2018recent}. The ${}^{239}$Pu nucleus has nuclear spin $I=\frac{1}{2}$ and is of great interest as an on-site probe of the rich $f$-electron physics of Pu. The first attempt to observe ${}^{239}$Pu NMR was performed on $\alpha$-Pu more than 50 years ago~\cite{Butterworth:1958hz}, however to date there is only a single report of ${}^{239}$Pu NMR~\cite{Yasuoka:2012fc} in the ionic insulator PuO$_2$. The main difficulty involved in observing ${}^{239}$Pu NMR---and other $f$-electron nuclei, in general---can be traced to the very strong hyperfine fields at the nucleus produced by on-site hyperfine coupling to the $f$-electrons. Consequently, the resulting spectral width can be very large, and the spin-lattice ($T_1$) and spin-spin ($T_2$) relaxation times can be extremely short, which makes detection of the signal difficult. These effects can be minimized in systems with a gap in the electronic and spin excitation spectrum, as evident in the case of PuO$_2$, UO$_2$, and YbB$_{12}$~\cite{Yasuoka:2012fc,Ikushima:2001jj,Ikushima:2000fc}.

Here we report the observation of, and the microscopic properties extracted from ${}^{239}$Pu NMR in powdered and single crystalline PuB$_4$. Crystals were grown by an Al-flux method  and sample preparation details are provided in the Supplementary Material. We deduce the resonant condition of ${}^{239}$Pu in PuB$_4$ ${}^{239}\gamma(1+K_{\mathrm{PuB}4})/2\pi = 2.288 \pm 0.001$~MHz/T from the powder spectra, and find axial symmetry of the hyperfine interaction on the Pu site. Both the powder and the single crystal Knight shift $K(T)$ of ${}^{239}$Pu show temperature dependence consistent with gap-like behavior with a static energy gap (extracted from the single crystalline $K_c(T)$ data) $\Delta_{K} \approx 156$~meV. The relaxation time is quite long---on the order of milliseconds to seconds---even at the ${}^{239}$Pu site, indicating that the $f$-electron configuration is non-magnetic. The dominant temperature dependence of the spin-lattice relaxation rate $T_1^{-1}(T)$ also shows gap-like behavior with a dominant dynamic gap $\Delta_{T_1} \approx 251$ meV. We compare our experimental NMR results with the density of states, calculated within density functional theory including spin-orbit coupling, which finds a gap of similar order of magnitude. A weak low-temperature peak in $T_1^{-1}(T)$ indicates the presence of bulk in-gap magnetic states with a gap $\delta \approx 2$ meV.

Our DFT calculations including spin-orbit coupling reveal a gap in the density of states (DOS) at the Fermi energy $E_F$ of roughly 254 meV as shown in Fig.~\ref{fig:PuB4_unit_cell_DOS}(b). To account for the presence of correlations we also performed DFT + DMFT calculations. Using a $U$ of 4.5 eV and high-order Slater integrals amounting to an effective $J=0.512$ eV~\cite{Yee:2010vf,Zhu:2013ss} and attempting to stabilize a magnetic solution, we find that the self-consistent solution recovers a non-magnetic state with a band gap at the Fermi level of order 10.3 meV (see Supplemental Material for further calculation details). The appreciable calculated gap in the DOS combined with an expected non-magnetic ground state indicate the probable absence of strong spin- and charge-relaxation channels, and therefore, we expect the spin-lattice relaxation rate in PuB$_4$ to be long enough to observe the ${}^{239}$Pu signal. The ${}^{239}$Pu nucleus has $I=\frac{1}{2}$ and the bare gyromagnetic ratio was determined based on the initial observation in PuO$_2$ to be ${}^{239}\gamma/2\pi = 2.29 \pm 0.001$~MHz/T~\cite{Yasuoka:2012fc}. Consequently, we would expect to find an NMR signal in the field range of roughly 7 to 9 T with an rf excitation frequency $f_0 \sim 20$ MHz. Indeed, for $f_0 = 20.222$ MHz we discovered an asymmetric powder spectrum between 8.80 and 8.92 T as shown in Fig.~\ref{fig:Pu239_spectra_layout}(a-b). To establish that the observed signal is indeed due to ${}^{239}$Pu from PuB$_4$ field-swept spectra were collected at several frequencies. These spectra are shown in Fig.~\ref{fig:Pu239_spectra_layout}(a) and they confirm the intrinsic nature of the NMR signal.

The crystal structure of PuB$_4$ has a single Pu site with oriented site symmetry \textit{m.2m} (see Fig.~\ref{fig:PuB4_unit_cell_DOS}(a)). For each crystallite in the powdered sample the resonance condition can be expressed as 2$\pi f_0$ = $\gamma B_0(1 + K_i)$ where $K_i$ are the elements of the shift tensor for a given field orientation and $B_0$ is the magnetic field at which the resonance occurs for frequency $f_0$. Although the local symmetry is orthorhombic in principle, the non-axial components of the shift tensor are found to be extremely close to zero from the spectral pattern in Fig.~\ref{fig:Pu239_spectra_layout}(b), \textit{i.e.}, it can be practically regarded to be tetragonal. Assuming tetragonal symmetry for the hyperfine interaction on Pu, the isotropic and axial shifts ($K_{iso}$ and $K_{ax}$, respectively) are extracted from the observed $K_c$ and $K_{ab}$ using $K_{iso} = (K_c + 2K_{ab})/3$ and $K_{ax} = (K_c - K_{ab})/3$, where the angular dependence of the shift is given by $K(\theta) = K_{iso} + K_{ax} (3\cos^2{\theta} - 1)$. 

The isotropic shift of ${}^{239}$Pu in PuB$_4$ is $K_{iso}(T = 4\mathrm{~K}) = -0.09 \pm 0.04$~\% is obtained from the slope in the frequency vs. field plot in Fig.~\ref{fig:Pu239_spectra_layout}(a). This value is notably different from the shift $K(T = 4\mathrm{~K}) = 24.72 \pm 0.04$~\% of ${}^{239}$Pu in PuO$_2$~\cite{Yasuoka:2012fc}. To calculate these shifts we have assumed the bare ${}^{239}\gamma/2\pi = 2.29$ MHz/T as determined from the study of PuO$_2$~\cite{Yasuoka:2012fc}. $K_{ax}(T = 4 \mathrm{~K}) = -0.48 \pm 0.01$~\% is also significantly different from the shift found in PuO$_2$~\cite{Yasuoka:2012fc} at the same temperature. It is worth noting that the relatively small absolute value of $K_{ax}$ was crucial to find the ${}^{239}$Pu signal in an anisotropic environment. 

The temperature dependence of the field-swept spectra at $f_0 = 20.222$ MHz and the corresponding least-squares fits are shown in Fig.~\ref{fig:Pu239_spectra_layout}(b). An axially symmetric shift tensor remains a good approximation for all temperatures measured. Fig.~\ref{fig:Pu239_spectra_layout}(c) illustrates that $K_{iso}$ has a small negative value with a positive temperature dependence, and $K_{ax}$ has a larger negative value with a smaller temperature dependence relative to $K_{iso}$. In general, $K_{iso}$ originates from the spin-polarized Fermi contact interaction and couples to the uniform spin susceptibility via the hyperfine interaction. $K_{ax}$ may be dominantly attributed to the temperature independent orbital hyperfine interaction with a small temperature dependence resulting from a reduction of the anisotropy of the spin susceptibility with increasing temperature. The facts that the spin-lattice relaxation time in PuB$_4$ is sufficiently long to enable the observation of ${}^{239}$Pu NMR, and that Knight shifts are weakly temperature dependent imply that the electronic state of Pu in PuB$_4$ is nearly nonmagnetic. Assuming a local picture this implies either that Pu has a $5f^6$ configuration or PuB$_4$ adopts a Kondo insulating state.

\begin{figure}[h!] 
	\includegraphics[trim=0cm 0cm 0cm 0cm, clip=true, width=\linewidth]{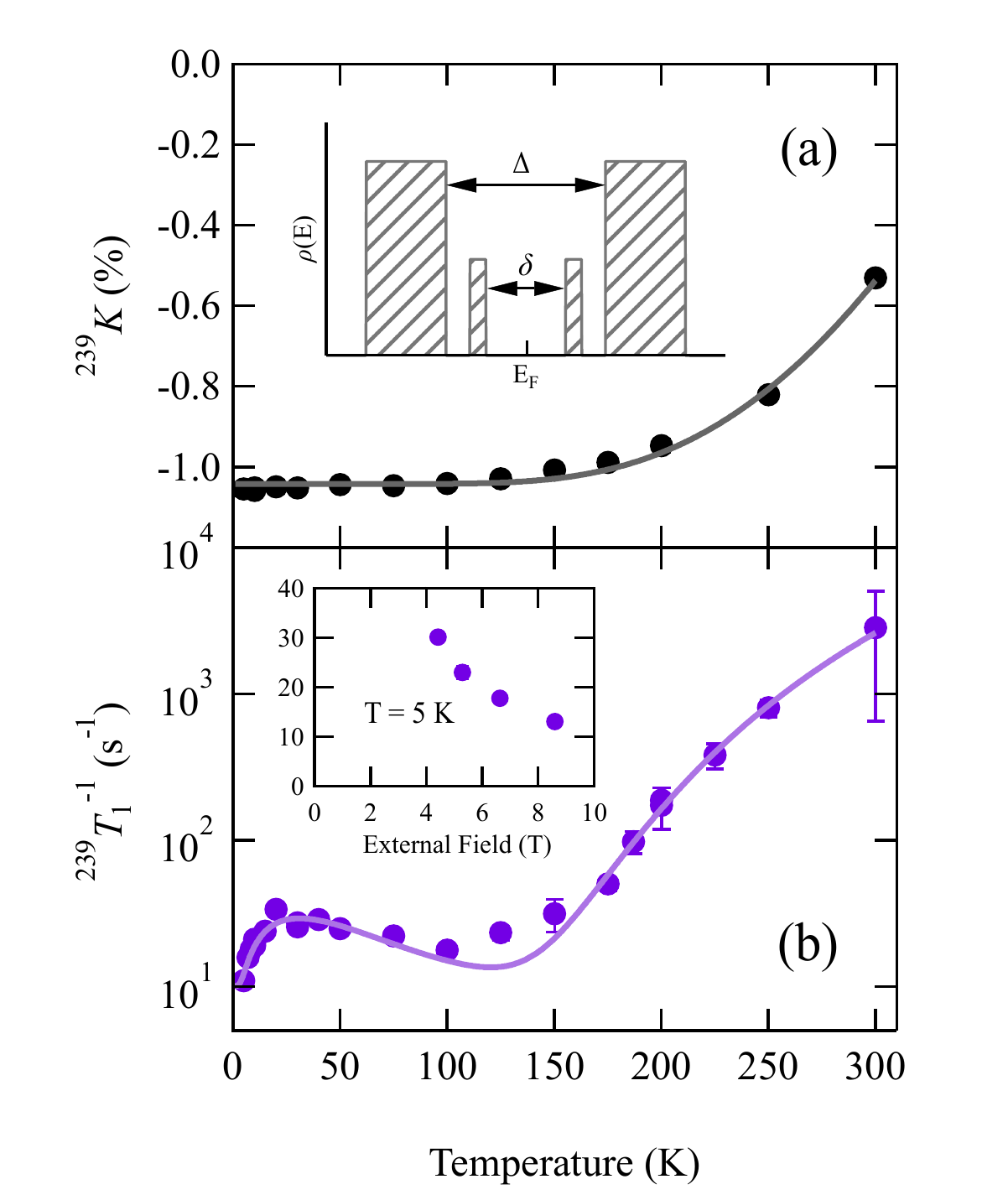}
	\caption{\label{fig:Pu239_sx_K_T1inv_for_PRL_inset}Single crystal ${}^{239}$Pu NMR data for external field aligned along the crystalline $\hat{c}$ direction. (a) Shift $K_c$ vs. temperature and fit (solid line) to gap-like behavior as discussed in the text which yields an energy gap $\Delta_{K} = 155.6 \pm 11.0$ meV, but is not sensitive to the in-gap states. The inset shows the model density of states $\rho(E)$ vs. energy $E$ employed in the fits. (b) Spin-lattice relaxation rate $T_1^{-1}$ vs. temperature and fit (solid line) to gap-like behavior as discussed in the text which yields energy gaps $\Delta_{T_1} = 251.3 \pm 49.4$ meV and $\delta = 1.8 \pm 2.4$ meV. The external field was adjusted (from 8.5900 T at 5 K to 8.5455 T at 300 K) such that the observed frequency was $f_0 = 19.465$ MHz for all temperatures. The inset shows the field dependence of $T_1^{-1}$ at $T= 5$ K indicating the suppression of in-gap states with applied external field.}
\end{figure}

Finally, we performed measurements on a single crystal of PuB$_4$ for the external field applied along the $\hat{c}$-axis. We measured both the $\hat{c}$-axis ${}^{239}$Pu shift $K_c$ and $T_1^{-1}$ as a function of temperature up to 300 K as shown in Fig.~\ref{fig:Pu239_sx_K_T1inv_for_PRL_inset}. We fit the ${}^{239}$Pu inversion recovery curves to the form 
\begin{equation}
\label{eqn:I_1_over_2_strexp}
	M_N(t) = M_N(\infty)\left(1 - {\alpha}e^{-(t/T_1)^\beta}\right),
\end{equation}
where $M_N(\infty)$ is the equilibrium nuclear magnetization, $\alpha$ is the inversion fraction, $T_1$ is the spin-lattice relaxation time, and $\beta$ is a stretching exponent that modifies the expected single exponential behavior ($\beta = 1$). We find that $\beta_{avg}=0.813$, which is a measure of the width of the probability distribution of $T_1$~~\cite{Johnston:2006st}, is independent of temperature and may indicate sensitivity to self-irradiation induced disorder~\cite{Booth:2013si}.  Both  $K_c$ and $T_1^{-1}$ are consistent with gap-like behavior, and $T_1^{-1}$ exhibits a low temperature maximum consistent with the presence of in-gap states which are suppressed with applied magnetic field as shown in the inset of Fig.~\ref{fig:Pu239_sx_K_T1inv_for_PRL_inset}.

From a chemistry perspective, the ${}^{239}$Pu orbital shift is very different between PuO$_2$ ($\sim +24.7$ \%~\cite{Yasuoka:2012fc}) and PuB$_4$ ($\sim -0.5$ \%). The origin of the difference in magnitude of the orbital shift is clear from the fact that in the case of PuO$_2$ the Pu ion has a completely ionic Pu$^{4+}$ ($5f^4$) state and experiences strong cubic crystalline electronic field giving rise to a non-magnetic ground state with a Van Vleck orbital magnetism, which is the main source of the hyperfine interaction to the Pu nuclear moment. In contrast, DFT + DMFT calculations point to PuB$_4$ being a strongly correlated insulator with possible strong topological character, similar to the case of SmB$_6$. In SmB$_6$ the gap arises from hybridization between $4f$ and ligand electrons that give rise to a pronounced non-integral value of the $4f$ configuration. Our results suggest that this is also the case in PuB$_4$. The large difference in orbital shift between PuO$_2$ and PuB$_4$ clearly indicates that ${}^{239}$Pu NMR is highly sensitive to the degree of bond mixing and the $f$-electron configuration. Furthermore, the relaxation time is roughly two orders of magnitude shorter than in PuO$_2$~\cite{Yasuoka:2012fc}, which likely reflects the difference in chemical environments between PuB$_4$ and PuO$_2$.

The capability to measure ${}^{239}$Pu was key to observing gap-like behavior in the static and dynamic spin-susceptibilities as evidenced by the temperature dependencies of $K_c$ and $T_1^{-1}$ shown in Fig.~\ref{fig:Pu239_sx_K_T1inv_for_PRL_inset}. Our ${}^{11}$B measurements of the temperature dependence of the Knight shift (see Supplemental Material) do not show any evidence of gap-like behavior, likely due to the much smaller value of the hyperfine coupling of the ${}^{11}$B nuclei to the electrons as compared to the ${}^{239}$Pu hyperfine coupling, which is expected to be on the order of 150 T/$\mu_B$. Therefore, our ${}^{239}$Pu NMR results are sensitive to otherwise enigmatic physics in PuB$_4$.

There exist a number of previous NMR studies that find gap-like behavior of $f$-electron systems, \textit{e.g.} SmB$_6$~\cite{Takigawa:1981sm, Caldwell:2007sm}, YbB$_{12}$~\cite{Ikushima:2000fc}, Ce$_3$Bi$_4$Pt$_3$~\cite{Reyes:1994bi}. Here we follow the analysis scheme of SmB$_6$~\cite{Caldwell:2007sm} by fitting the temperature dependence of the ${}^{239}$Pu Knight shift and spin-lattice relaxation rate by assuming a simple model for the density of states near the Fermi energy. The Knight shift is given by,
\begin{equation}
\label{eqn:K_DOS_fermi}
	K(T) \propto \int{f(E,T)[1 - f(E,T)] \rho(E) dE},
\end{equation}
where $f(E,T)$ is the Fermi function and $\rho(E)$ is the density of states. The spin-lattice relaxation rate is given by,
\begin{equation}
\label{eqn:T1inv_DOS_fermi}
	T_1^{-1}(T) \propto \int{f(E,T)[1 - f(E,T)] \rho(E)^2 dE}.
\end{equation}
We assume a simplified model of the density of states (equivalent to that of Caldwell \textit{et al.}~\citep{Caldwell:2007sm}) given by,
\begin{equation}
\label{eqn:DOS_hyb_gap}
\begin{split}
	\rho(E) &= \rho_i(T) \mathrm{~for~} \delta < |E| < W_i\\
	    		&= \rho \mathrm{~for~} \Delta < |E| < W,
\end{split}
\end{equation}
and zero otherwise as shown in the inset of Fig.~\ref{fig:Pu239_sx_K_T1inv_for_PRL_inset}(a). We perform least-squares fits using a Levenberg-Marquardt minimization algorithm which iteratively recalculates the model function via numerical integration of Eqns.~\ref{eqn:K_DOS_fermi} and \ref{eqn:T1inv_DOS_fermi} (see Supplemental Material for a full description of the curve fitting). The energy gap extracted from the Knight shift $\Delta_{K} = 155.6 \pm 11.0$. For the Knight shift we find no indication of the presence of in-gap states, that is $\rho_i(T) = 0$, similar to the static susceptibility of SmB$_6~$\cite{Caldwell:2007sm}. The dominant energy gap extracted from the spin-lattice relaxation $\Delta_{T_1} = 251.3 \pm 49.4$ meV. A smaller in-gap density of states $\rho_i(T) = \rho_{i0} e^{-T/T_0}$ with an energy gap $\delta = 1.8 \pm 2.4$ meV was also found to be consistent with the small low temperature enhancement of $T_1^{-1}(T)$. The discrepancy between the static gap $\Delta_{K}$ and the dynamic gap $\Delta_{T_1}$ has been observed in numerous spin-gap systems~\cite{Itoh1997spingaps} and is related to differences in the processes that contribute to the Knight shift and the spin-lattice relaxation. In the majority of these spin-gap systems the dynamic gap $\Delta_{T_1} > \Delta_{K}$ and on average $\Delta_{T_1}/ \Delta_{K} = 1.73$. In the case of PuB$_4$ we find $\Delta_{T_1}/ \Delta_{K} = 1.6 \pm 0.3$.

While NMR is not sensitive to the surface states in bulk powders or single crystals~\cite{Koumoulis:2013ti}, it is a powerful microscopic probe of the bulk properties of topological materials. Our ${}^{239}$Pu NMR results are consistent with a  bulk gap which is only slightly suppressed from the DFT+SOC calculated value of 254 meV. In addition to the dominant gap-like behavior evidenced by $K_c(T)$ and $T_1^{-1}(T)$, we also find a small peak at low temperature that is reminiscent of the ${}^{11}$B $T_1^{-1}(T)$ in SmB$_6$~\cite{Takigawa:1981sm} and YbB$_{12}$~\cite{Ikushima:2000fc}. In SmB$_6$ the peak is thought to be due to bulk magnetic in-gap states, and while the nature of these states is still controversial, it has been suggested that these states are identical to the topologically protected surface states~\cite{Tetsuya:2011sm}. In PuB$_4$ we find that $T_1^{-1}(T=5\mathrm{~K})$ is strongly field dependent as shown in the inset of Fig.~\ref{fig:Pu239_sx_K_T1inv_for_PRL_inset}(b), which is similar to previous field dependent measurements of SmB$_6$~\cite{Caldwell:2007sm}. 

These results motivate further investigation of the field and Pu-substitution dependence of $T_1^{-1}$ over a wide temperature range. Previous transport measurements find a much smaller gap $\Delta = 35$ meV~\cite{Choi:2018up}. This is also the case in YbB$_{12}$, where NMR finds a larger gap than resistivity, and may be related to the presence of in-gap states which account for the low temperature enhancement in $T_1^{-1}$. This discrepancy motivates Hall coefficient measurements in PuB$_4$ (which in YbB$_{12}$ agree with the NMR-measured gap), as well as surface-sensitive tunneling or spin-polarized ARPES measurements. Finally, we note that measurements comparing ${}^{11}$B and ${}^{10}$B $T_1^{-1}$ in YbB$_{12}$ and Yb$_{0.99}$Lu$_{0.01}$B$_{12}$ provide evidence for another interpretation of the low temperature relaxation enhancement, namely that it may be driven by fluctuations of defect-induced magnetic centers and spin-diffusion-assisted relaxation~\cite{Shishiuchi2002YbB12_spin_diff}. These YbB$_{12}$ results motivate further measurements and comparison of ${}^{11}$B and ${}^{10}$B $T_1^{-1}$ in PuB$_4$.

To conclude, we have performed ${}^{239}$Pu NMR measurements for the second time ever in powdered and single crystalline PuB$_4$. We extracted the isotropic and anisotropic shifts from the uniaxially symmetric powder pattern and demonstrate that one can observe the ${}^{239}$Pu NMR signal in anisotropic environments and up to room temperature. The large contrast of the orbital shift between the purely ionic insulator PuO$_2$ ($\sim +24.7$ \%) and  band insulator PuB$_4$ ($\sim -0.5$ \%) provide us with new tool to investigate the nature of the chemical bond based on the value of the ${}^{239}$Pu shift. Single crystal ${}^{239}$Pu NMR measurements of $K_c(T)$ and $T_1^{-1}(T)$ provide unique access to bulk gap-like behavior with an energy gap that is only slightly suppressed with respect to DFT+SOC calculations, and $T_1^{-1}(T)$ also evidences the existence of bulk in-gap states. Our confirmation of a bulk gap motivate future surface sensitive measurements to confirm the theoretical prediction that PuB$_4$ is a topological insulator.

\begin{acknowledgments}
\section{Acknowledgments}
The authors would like to thank D. L. Clark, Z. Fisk, P. F. S. Rosa, A. M. Mounce, S. Seo, R. Movshovich, M. Janoschek, D.-Y. Kim, D. Fobes, N. Sung, N. Leon-Brito, M. W. Malone, H.-J. Grafe, M. Po\v{z}ek, D. Kasinathan and P. Coleman for stimulating discussions. Work at Los Alamos National Laboratory was performed with the support of the Los Alamos LDRD program. TEA-S was supported as part of the Center for Actinide Science and Technology (CAST), an Energy Frontier Research Center funded by the U.S. Department of Energy, Office of Science, Basic Energy Sciences under Award Number DE-SC0016568. HS was also partly supported by JSPS KAKENHI Grant Number JP16KK0106. APD acknowledges a Director's Postdoctoral Fellowship supported through the Los Alamos LDRD program.
\end{acknowledgments}

\bibliography{PuB4_NMR}

\end{document}


\title{${}^{239}$Pu nuclear magnetic resonance in the candidate topological insulator PuB$_4$ - Supplemental Material}

\author{A. P. Dioguardi}
\email{adioguardi@gmail.com}
\affiliation{Los Alamos National Laboratory, Los Alamos, New Mexico 87545, USA}
\affiliation{IFW Dresden, Institute for Solid State Research, P.O. Box 270116, D-01171 Dresden, Germany}

\author{H. Yasuoka}
\affiliation{Los Alamos National Laboratory, Los Alamos, New Mexico 87545, USA}
\affiliation{Max Planck Institute for Chemical Physics of Solids, 01187 Dresden, Germany}

\author{S. M. Thomas}
\affiliation{Los Alamos National Laboratory, Los Alamos, New Mexico 87545, USA}

\author{H. Sakai}
\affiliation{Advanced Science Research Center, Japan Atomic Energy Agency, Tokai, Naka, Ibaraki 319-1195, Japan}
\affiliation{Los Alamos National Laboratory, Los Alamos, New Mexico 87545, USA}

\author{S. K. Cary}
\author{S. A. Kozimor}
\affiliation{Los Alamos National Laboratory, Los Alamos, New Mexico 87545, USA}

\author{T. E. Albrecht-Schmitt}
\affiliation{Department of Chemistry and Biochemistry, Florida State University, 95 Chieftan Way, Tallahassee, Florida 32306}

\author{H. C. Choi}
\author{J.-X. Zhu}
\author{J. D. Thompson}
\author{E. D. Bauer}
\author{F. Ronning}
\affiliation{Los Alamos National Laboratory, Los Alamos, New Mexico 87545, USA}

\date{\today}

\maketitle

\section{Experimental and theoretical methods}

Single crystals of PuB$_4$ were grown from aluminum flux.  Elemental $\alpha$-Pu (isotopic mixture: 0.013 wt. \% ${}^{238}$Pu, 93.96\% ${}^{239}$Pu, 5.908\% ${}^{240}$Pu, 0.098\% ${}^{241}$Pu, and 0.025\% ${}^{242}$Pu) was placed in a 10 ml alumina crucible along with boron pieces (99.9999\%) and aluminum shot (99.999\%) in the molar ratio Pu:B:Al=1:6:600. The constituents were heated in flowing purified argon to 1450 ${}^{\circ}$C at a rate of 50 ${}^{\circ}$C/hour, held at 1450 ${}^{\circ}$C for 10 hours, then cooled slowly at 5 ${}^{\circ}$C/hour to 1000 ${}^{\circ}$C at which point the furnace was turned off. After the furnace reached room temperature the crucibles were removed from the furnace.  The aluminum was dissolved from the crucible by etching with a NaOH (500 mL, 8 M) solution. The etching was preformed inside a 1 L Pyrex beaker, which was capped with Teflon foil, all of which sat in an 1170 mL Pyrex crystallization dish. After two days all of the aluminum dissolved  leaving behind well-faceted PuB$_4$ crystals with typical dimensions 1$\times$1$\times$0.3 mm$^3$. No trace of the cubic PuB$_6$ structure was found in the sample batch. The single crystal X-ray diffraction results indicate full occupancy of the B sites with no interstitial or free B.

The PuB$_4$ single crystals were ground to a powder using an agate mortar and pestle. The powder was then sifted through a 125-$\mu$m sieve, then through a 30-$\mu$m sieve to remove the finest particles. 125 mg of powder between 30 and 125 $\mu$m was loaded into an NMR coil embedded in a Stycast 1266 epoxy cube with dimensions 20$\times$20$\times$20 mm$^3$ to avoid radioactive contamination. The cube was hollow along the coil's axis with both of its ends sealed by titanium frits with 2-$\mu$m diameter pores, allowing for thermal contact via He gas heat transfer.

All NMR spectral data were collected by performing an optimized spin-echo $\pi/2 - \tau - \pi$ pulse sequence. $T_1^{-1}$ was measured by integration of the phase-corrected real part of the spin echo following an inversion or saturation recovery $\pi - t_{wait} - \pi/2 - \tau - \pi$ pulse sequence. 
Home-built NMR probes with variable cryogenic capacitors were used in conjunction with superconducting NMR magnets that have a field homogeneity better than 10 ppm over a 1-cm diameter spherical volume.

DFT calculations were performed using the WIEN2k code~\cite{Schwarz:2003ss}. For the exchange-correlation potential we used the parameterization of Perdew-Burke-Ernzerhof based on the generalized gradient approximation~\cite{Perdew:1996gg}, and spin-orbit interactions were included through a second variational method. The electric field gradient is subsequently determined directly as a second derivative of the electrostatic potential, which is obtained by solving Poisson's equation using the self-consistently determined charge density from the all-electron DFT calculation~\cite{Blaha:1985fp}. The effects of correlations were included using a DFT+DMFT approach~\cite{Kotliar:2006es,Haule:2010dy}. Allowing a spin-polarized solution for the DFT calculation incorrectly leads to a metallic solution with a magnetic moment of 4.4 $\mu_\text{B}$ per formula unit.

Fully self-consistency calculations in both charge density and impurity Green's functions were carried out at a temperature of $T=116$ K. The Coulomb interaction $U=4.5$ eV was used together with remaining Slater integrals $F^2=6.1$ eV, $F^4=4.1$ eV and $F^6=3.0$ eV reduced by 30\%. The quantum impurity model was solved by using the strong coupling version of the continuous time quantum Monte Carlo method~\cite{Werner:2006he,Haule:2007qm}. The calculations involved at least 10 iterations of charge self-consistency (each containing 1 DMFT iteration and 10 LDA iterations).

\section{Control experiment}
\label{sec:control_experiment}

As ${}^{239}$Pu nuclear magnetic resonance (NMR) has only been observed in one previous compound, it was prudent to perform a control experiment to confirm that the observed signal was indeed intrinsic to the PuB$_4$. The same experiment was conducted after replacing the sample and encapsulation cube with an identical but empty cube. The rest of the experimental apparatus was unchanged and operated under identical conditions of temperature, magnetic field, and observation frequency. No NMR signal was observed in this control experiment. Furthermore, we performed a two-dimensional rf pulse power optimization experiment on the empty control-cube at the resonance conditions stated in the main text. No signal was observed for any excitation conditions.

\section{${}^{239}$Pu hyperfine coupling analysis}
\label{sec:hyperfine_coupling}

\begin{figure} 
	\includegraphics[trim=0cm 0cm 0cm 0cm, clip=true, width=0.9\linewidth]{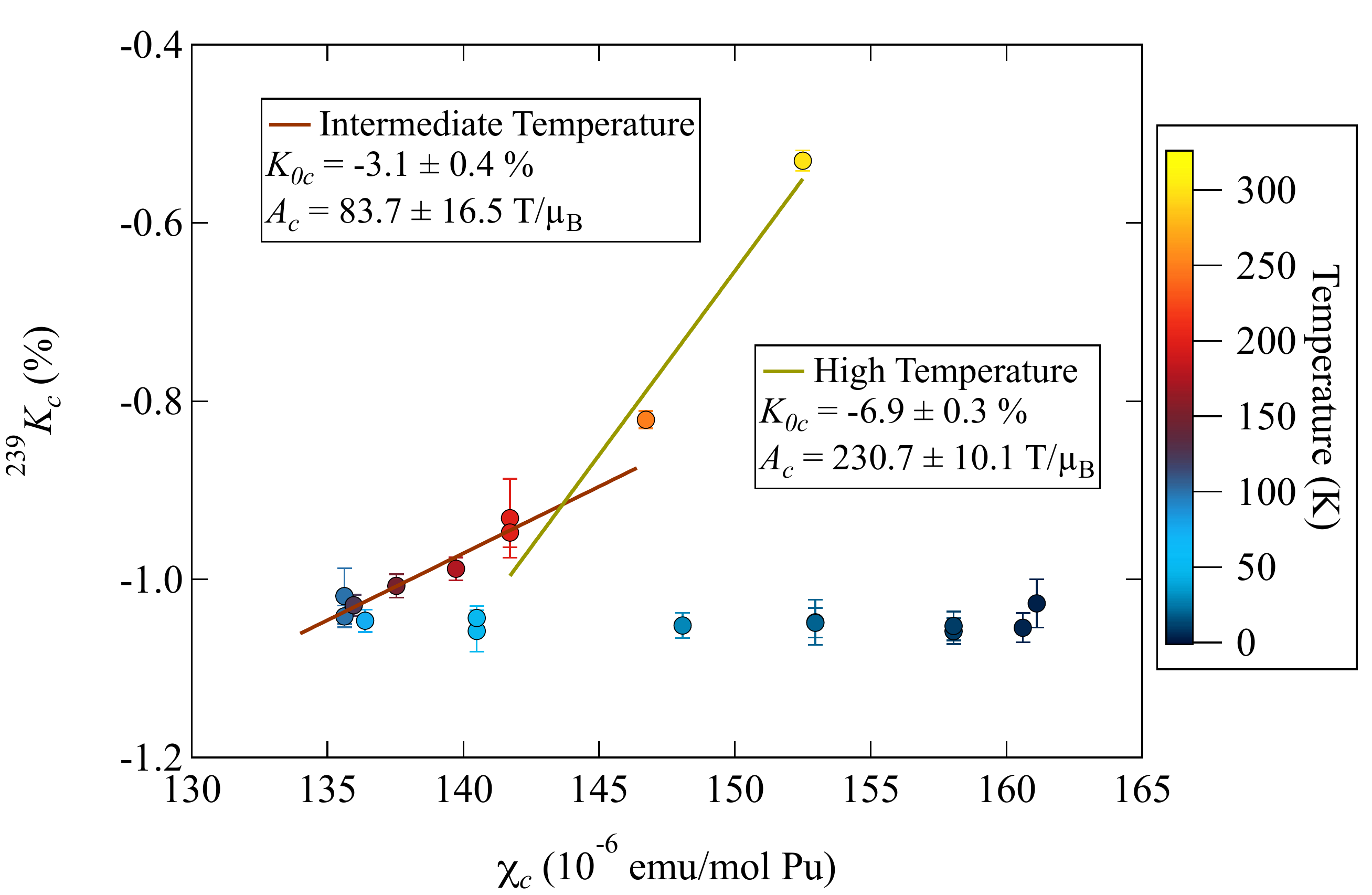}
	\caption{\label{fig:239Kc_vs_chic_sx} Knight shift $K_c$ for external field aligned along the crystalline $c$-axis vs the bulk magnetic susceptibility $\chi_c$ for the same crystal orientation with temperature as an implicit parameter.}
\end{figure}

We attempt a naive analysis of the ${}^{239}$Pu hyperfine coupling $A_c$ in a single crystal sample of PuB$_4$ for external field aligned with the crystalline $c$-axis by plotting $K_c$ vs $\chi_c$ with temperature as an implicit parameter. We observe a positive slope of the $K$-$\chi$ plot and use the slope of a linear fit to estimate the hyperfine coupling $A_c$. The $K$-$\chi$ plot is not linear, and therefore we performed fits to two separate regions. The high temperature region was taken to be 200-300 K and the intermediate temperature region was chosen to be 100-200 K. The hyperfine couplings quoted in Fig.~\ref{fig:239Kc_vs_chic_sx} should be taken with caution as it is known that the relationship between the hyperfine coupling and the slope of the $K$-$\chi$ plot is no longer trivial in the case of strong spin-orbit coupling~\cite{nisson2016}. In spite of this caveat we find a relatively large hyperfine coupling at high temperatures on the order of the free-ion calculation of 258.3 T/$\mu_\mathrm{B}$ for Pu${}^{4+}$~\cite{Yasuoka:2012fc}. Below 100 K the Knight shift saturates and remains constant, whereas the susceptibility increases down to the lowest temperatures. This low temperature tail in the bulk susceptibility likely results from paramagnetic impurities. The Knight shift is unaffected by the small concentration of impurities and is able to probe the small of density of states within the gap. At high temperatures the susceptibility increases only slightly, likely due to the nearly nonmagnetic state of the Pu $f$-electrons in PuB$_4$.

\section{${}^{239}$Pu gap fitting procedure}
\label{sec:239Pu_gap_fitting_procedure}

In the main text we briefly describe the fitting procedure for extracting the gap values of the static gap $\Delta_k$ and the dynamic gap $\Delta_{T_1}$. Here we detail the exact method and all optimized fit parameters for the benefit of the reader (with some repetition for clarity) .The general form of the Knight shift is given by,
\begin{equation}
\label{eqn:K_DOS_fermi}
	K(T) \propto \int{f(E,T)[1 - f(E,T)] \rho(E) dE},
\end{equation}
where $f(E,T)$ is the Fermi function and $\rho(E)$ is the density of states (DOS). The spin-lattice relaxation rate is given by,
\begin{equation}
\label{eqn:T1inv_DOS_fermi}
	T_1^{-1}(T) \propto \int{f(E,T)[1 - f(E,T)] \rho(E)^2 dE}.
\end{equation}
In the above equations the Fermi function given by,
\begin{equation}
\label{eqn:fermi_dirac_distribution}
	f(E,T) = \frac{1}{\exp{\left(\frac{E - E_F}{2k_BT}\right)} + 1 },
\end{equation}
where $E_F$ is the Fermi energy and $k_B$ is the Boltzmann constant. We model the DOS as,
\begin{equation}
\label{eqn:DOS_hyb_gap}
\begin{split}
	\rho(E) &= \rho_i(T) \mathrm{~for~} \delta < |E| < W_i\\
	    		&= \rho \mathrm{~for~} \Delta < |E| < W,
\end{split}
\end{equation}
and zero otherwise. Within this model the in-gap DOS must have some temperature dependence due to the fact that the relaxation rate is peak-like and not simply activated. Following the same analysis scheme used by Caldwell \textit{et al.} for SmB$_6$~\cite{Caldwell:2007sm}, we express the in-gap DOS as
\begin{equation}
\label{eqn:DOS_in_gap_temp_dep}
    \rho_i(T) = \rho_{i0} e^{-T/T_0},
\end{equation}
where $\rho_{i0}$ is the maximum value of the in-gap DOS and $T_0$ is a characteristic temperature scale. For the actual fits we must include a temperature independent offset $K_0$ for the case of $K(T)$ and $(T_1^{-1})_0$ is a temperature-independent residual relaxation rate.

\begin{table*}[t]
\centering
\begin{tabular}{cccccccccc}
              & $\Delta$ (meV)   & $\delta$ (meV) & $K_0$ (\%)           & $(T_1^{-1})_0$ (s$^{-1}$) & $\rho$ (arb. units) & $\rho_{i0}$ (arb. units) & $T_0$ (K) & $W$ (meV) & $W_i$ (meV) \\
\hline \\[-0.3cm]
$K(T)$	      & $155.6 \pm 11.0$ & - & $-1.042 \pm 0.005$ & - & $2.1 \pm 0.4$ & 0 & - & 1500 & - \\
$T_1^{-1}(T)$ & $251.3 \pm 49.4$ & $1.8 \pm 2.4$  & - & $10.1 \pm 7.2$	& $2560.3 \pm 1740.9$ & $169.1 \pm 22.9$ & $51.8 \pm 17.1$ & 1500 & 100
\end{tabular}
\caption{\label{tab:K_vs_T_gapfit_coefs}Fit coefficients for single crystal ${}^{239}$Pu $K(T)$ and $T_1^{-1}(T)$.}
\end{table*}

The fitting procedure was written in the Python programming language and makes use of the \verb|numpy| (numerical python) and \verb|lmfit| (Levenberg-Marquardt fit) packages. We write objective functions for $K(T)$(Eqn.~\ref{eqn:K_DOS_fermi}) and $T_1^{-1}(T)$ (Eqn.~\ref{eqn:T1inv_DOS_fermi}) which are integrated over energy for each temperature at which experimental data exists. We then input the respective objective function into a second minimization function that uses the \verb|lmfit| package's \verb|Minimizer| class. We hold the Fermi energy to be at the gap center as the gross features of the experimental data are not well reproduced for $E_F \not\approx 0$. Initially we explore the parameter space for physically reasonable parameter guesses for the free parameters ($rho_{i0}$, $T_0$, $\delta$, $W_i$, $\rho$, $\Delta$, $W$, and either $K_0$ or $(T_1^{-1})_0$). An initial minimization was performed using the  Nelder-Mead method which is fast and better at finding the global minimum of $\chi^2$, but is unable to calculate the standard errors. Then we use the minimized output fit coefficients of the Nelder-Mead method as an input  for the Levenberg-Marquardt $\chi^2$ minimizer, yielding the gap values and standard errors quoted in the main text of the manuscript and all coefficients shown in Table~\ref{tab:K_vs_T_gapfit_coefs}.

The fitting procedure for $K(T)$ is straightforward. Due to the fact that we do not observe any low temperature enhancement in the Knight shift, we hold the in-gap states to be zero by holding $rho_{i0} = 0$ and. We also verified that allowing $\rho_i(T)$ to vary results in $rho_{i0} \approx 0$. The bandwidth $W$ was found to evolve to values larger than 1 eV, and to have little effect on the functional form of the objective function for $W \gtrsim 1000$~meV and therefore we hold $W = 1500$~meV (this was also true for the $T_1^{-1}(T)$). So the only remaining parameters that were allowed to vary are $\rho$, $\Delta$, and $K_0$. The optimized values obtained for the parameters for $K(T)$ are given in the first row of Table~\ref{tab:K_vs_T_gapfit_coefs}. The numerical integral is carried out over the range $E = -3$ to $3$ eV with an energy resolution of 0.06 meV.

For $T_1^{-1}(T)$ the fitting procedure was more complicated due to the appearance of a low temperature peak similar to SmB$_6$ and YbB$_12$ for example. We hold the Fermi energy $E_F = 0$ and $W = 1500$~meV as mentioned above, and used the Nelder-Mead method to search the parameter space for reasonable initial guesses. We found that the bandwidth of the in-gap states had a similar lack of effect on the model function for $W_i \gtrsim 50$ meV, and so we chose to hold $W_i = 100$ meV. The best fit coefficients can be found in the second row of Table~\ref{tab:K_vs_T_gapfit_coefs}. We found good stability of the optimized parameters associated with the dominant gap-like behavior ($\rho$ and $\Delta$) independent of the initial guesses, indicating that we are indeed finding the global minimum of these coefficients in the parameter space. Even for the case where we ignore the in-gap states and hold $\rho_{i0} = 0$ we find $\Delta \sim 250$ meV. On the other hand, the best fit coefficients associated with the in-gap states have larger standard errors (\textit{e.g.} the small gap $\delta$). This is partially due to the small effect of the in-gap states on $T_1^{-1}(T)$, but also due to correlations between the parameters. As such, we do not draw conclusions in the main text from the exact values of these quantities.

\section{${}^{239}$Pu spin-lattice relaxation rate anisotropy}
\label{sec:239Pu_spin-lattice_relaxation_rate_anisotropy}

\begin{figure}[h!] 
	\includegraphics[trim=0cm 0cm 0cm 0cm, clip=true, width=\linewidth]{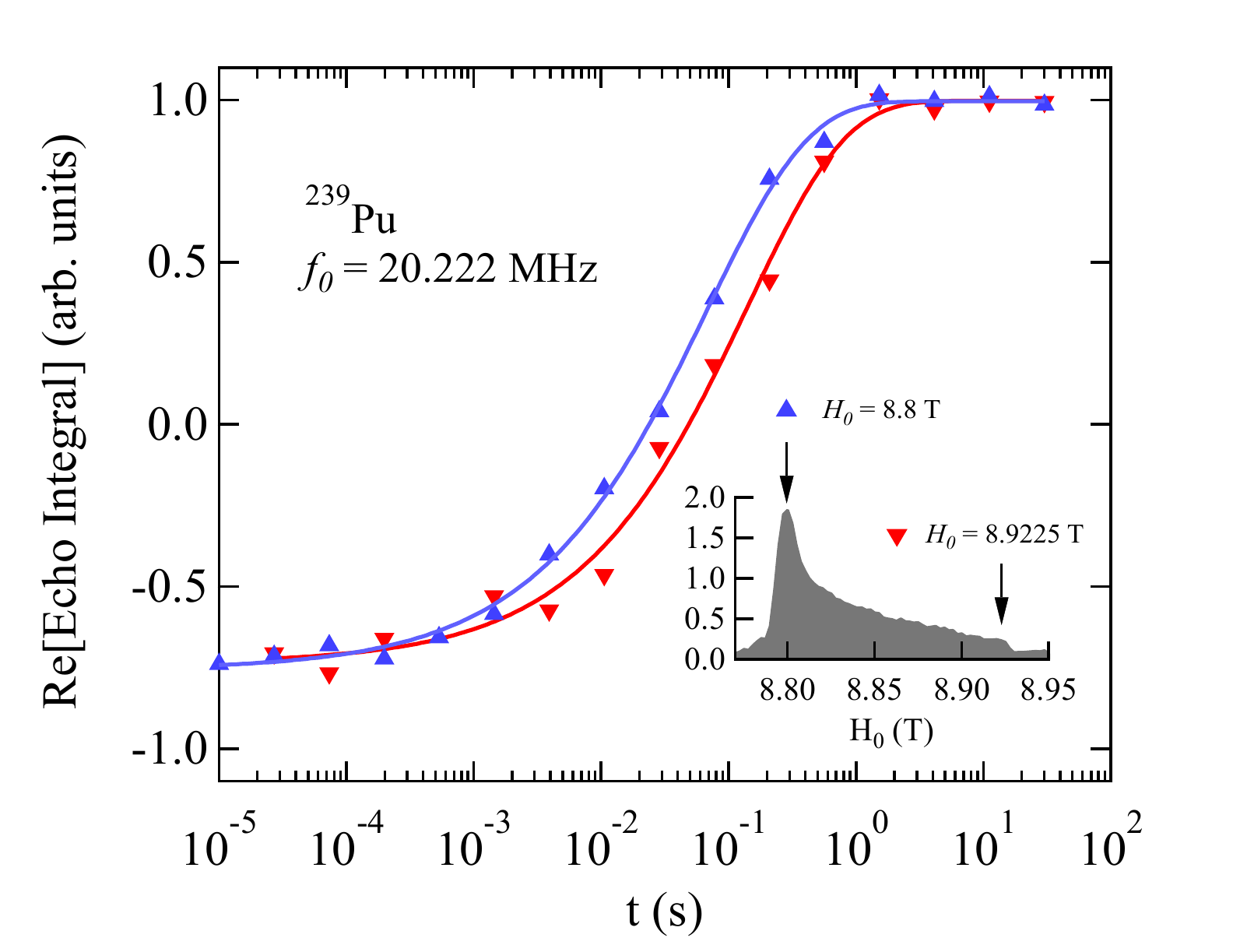}
	\caption{\label{fig:Pu239_T1_4K_aniso_recovery}Inversion recovery curve in powdered PuB$_4$ showing the real part of the integrated ${}^{239}$Pu spin-echo signal as a function of delay time $t$. The recovery curves are well fit by a stretched exponential function for an $I=\frac{1}{2}$ nucleus (Eqn.~\ref{eqn:I_1_over_2_strexp}) and show moderate anisotropy as discussed in the text. The inset shows the spectral positions where the recovery curves were acquired.}
\end{figure}

For an initial study of nuclear spin-lattice relaxation behavior in powdered single crystals of PuB$_4$, we measured the ${}^{239}$Pu nuclear magnetization recovery curves for both the $B_0 \parallel c$ shoulder of the spectrum (high-field side) and the $B_0 \perp c$ peak in a powder of PuB$_4$ (shown in Fig.~\ref{fig:Pu239_T1_4K_aniso_recovery}). We fit the ${}^{239}$Pu inversion recovery curves to the form 
\begin{equation}
\label{eqn:I_1_over_2_strexp}
	M_N(t) = M_N(\infty)\left(1 - {\alpha}e^{-(t/T_1)^\beta}\right),
\end{equation}
where $M_N(\infty)$ is the equilibrium nuclear magnetization, $\alpha$ is the inversion fraction, $T_1$ is the spin-lattice relaxation time, and $\beta$ is a stretching exponent that modifies the expected single exponential behavior ($\beta = 1$).

For the crystallites with $B_0 \perp c$ we find $T_1^{-1} = 14.81 \pm 0.93$ s$^{-1}$ and $\beta = 0.55 \pm 0.03$, and for crystallites with $B_0 \parallel c$ we find $T_1^{-1} = 7.16 \pm 92$ s$^{-1}$ and $\beta = 0.56 \pm 0.05$. In this case, $T_1^{-1}$ is the median of the probability distribution of relaxation rates and $\beta$ is a measure of the logarithmic full-width at half maximum, with the distribution approaching a Dirac delta function as $\beta \rightarrow 1$~\cite{Johnston:2006st}. The appearance of stretched exponential behavior is not completely clear, but is partially caused by a sensitivity to crystalline imperfections caused by grinding to produce the powdered sample. In our single crystal measurements discussed in the main manuscript we find that $\beta$ is much closer to unity and nearly temperature independent ($\beta_{avg}=0.813$); this may indicate sensitivity to self-irradiation induced disorder~\cite{Booth:2013si}. Even though we cannot extract precise values of the spin-lattice relaxation rates, we still glean useful information from the relaxation measurements. The median of the probability distribution for the ${}^{239}$Pu relaxation times is in the range of hundreds of milliseconds to seconds and agrees with the single crystal data.

\begin{table*}[t]
\centering
\begin{tabular}{cccccccc}
	&					&			&	 				& \multicolumn{2}{c}{Experimental results}			& \multicolumn{2}{c}{Theoretical calculations}	\\
Site	& Wyckoff Position	& Occupancy		& Shift (\%)						& $\lvert V_{zz} \rvert$ (MHz)	& $\eta$ (unitless)	& $\lvert V_{zz} \rvert$ (MHz) 	& $\eta$ (unitless)	\\
\hline \\[-0.3cm]
B(1)	& $4e$				& 1				& -0.0295						& 0.6							& 0					& 0.400                   	& 0		\\
B(2)	& $4h$				& 1				& -0.0295						& 0.42							& $<$ 0.03		& 0.579                		& 0.052 	\\
B(3)	& $8j$				& 2 				& -0.0525 					& 0.353						& 0.664			& 0.343                 		& 0.642	\\
\end{tabular}
\caption{\label{tab:EFG_parameters}${}^{11}$B experimental spectral parameters and DFT calculated EFGs.}
\end{table*}

\section{${}^{11}$B Powder NMR results}
\label{sec:11B_shift}

Data collected with ${}^{11}$B NMR in PuB$_4$ support our observations from ${}^{239}$Pu NMR. ${}^{11}$B is an excellent NMR nucleus with $I=\frac{3}{2}$ and ${}^{11}\gamma/2\pi = 13.6629814$ MHz/T~\cite{Harris:2001dv}. This is advantageous for the discovery of the ${}^{239}$Pu signal, as one can first optimize the pulse conditions using ${}^{11}$B NMR before searching for ${}^{239}$Pu NMR. Furthermore, it is useful to have a well-studied nucleus with which to compare the shift and therefore gain insight into the hyperfine coupling to the Pu $f$-electrons that dominate the magnetic properties of the material. A comparison can therefore be made between the on-site hyperfine coupling to the ${}^{239}$Pu and the transferred hyperfine coupling to the ${}^{11}$B.

\subsection{Spectral Measurements}
\label{ssec:11B_spectral_measurements}

\begin{figure*} 
	\includegraphics[trim=0cm 0cm 0cm 0cm, clip=true, width=0.9\linewidth]{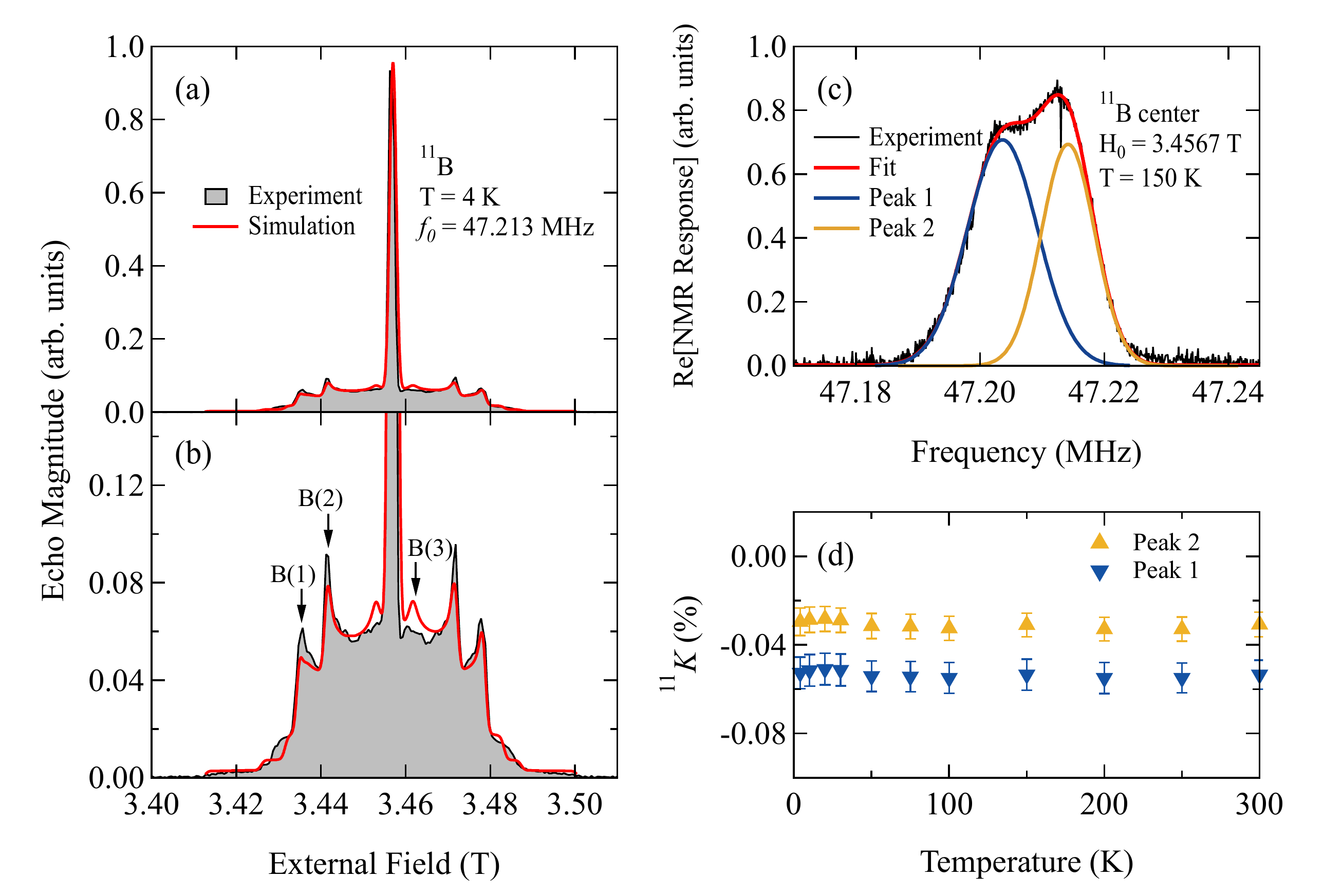}
	\caption{\label{fig:B11_spectra_layout}(a) and (b) ${}^{11}$B field-swept spectrum at $T = 4$ K with overlaid computer simulation shown as red lines. Experimental simulation parameters are shown in Table~\ref{tab:EFG_parameters}. (c) Multipeak fitting scheme employed for extraction of shift as a function of temperature from the two dominant central lines ${}^{11}$B. (d) Temperature dependence of the ${}^{11}$B shift ${}^{11}K$ of the two spectrally dominant overlapping central transitions determined from the two peaks shown in (c).}
\end{figure*}

There are three crystallographic B sites with local symmetries Tetragonal (\textit{4}), orthorhombic (\textit{m.2m}), and monoclinic (\textit{m}). As such, all B sites are allowed by symmetry to have a nonzero electric field gradient (EFG). This, coupled with the fact that ${}^{11}$B has $I = \frac{3}{2}$, results in a more complex powder spectrum than that of the ${}^{239}$Pu. The field-swept spectrum at $T = 4$ K is shown in Fig.~\ref{fig:B11_spectra_layout}(a) and (b). The spectrum consists of three overlapping but well-distinguished powder patterns with differing shifts and EFG parameters; the quantities used to generate the simulation in Fig.~\ref{fig:B11_spectra_layout} are listed in Table~\ref{tab:EFG_parameters}. We also compare the experimentally determined EFGs to those computed by DFT in Table~\ref{tab:EFG_parameters}. We note that there exists some ambiguity in the assignment of B(1) vs B(2) as the B(2) site has a very small experimental value of the EFG asymmetry parameter $\eta$.

To extract the temperature dependence of the ${}^{11}$B shift, we collected fast Fourier transforms of the sharp central transitions over a wide temperature range and employed a multipeak fitting scheme as shown in Fig.~\ref{fig:B11_spectra_layout}(c). We performed double-Gaussian fits on the two spectrally dominant central transitions of ${}^{11}$B (see Fig.~\ref{fig:B11_spectra_layout}). In this process we neglect the effect of second-order quadrupole perturbation, which will slightly shift and broaden the powder pattern by between 0\% (for $H_0 \parallel V_{zz}$, which is the principle axis of the EFG tensor) and 0.003\% (for $H_0 \perp V_{zz}$). The maximum second order perturbation of the central transition is roughly an order of magnitude less than the shifts that we find in Fig.~\ref{fig:B11_spectra_layout}(d). The extracted shifts for the two spectrally dominant peaks are shown as a function of temperature in Fig.~\ref{fig:B11_spectra_layout}(d). We expect a transferred hyperfine coupling to the Pu $f$-electrons mediated by B $p$-orbital hybridization, which can be quite small in comparison to the on-site hyperfine coupling of the ${}^{239}$Pu. We find little temperature dependence of the ${}^{11}$B Knight shift, which is in agreement with the bulk susceptibility~\cite{Smith:gj}. The temperature independence of the ${}^{11}$B Knight shift highlights the importance of the large on-site hyperfine coupling of ${}^{239}$Pu which allows for the observation of gap-like behavior in PuB$_4$.

\subsection{Spin-Lattice Relaxation}
\label{sec:slrr}

\begin{figure} 
	\includegraphics[trim=0cm 0cm 0cm 0cm, clip=true, width=\linewidth]{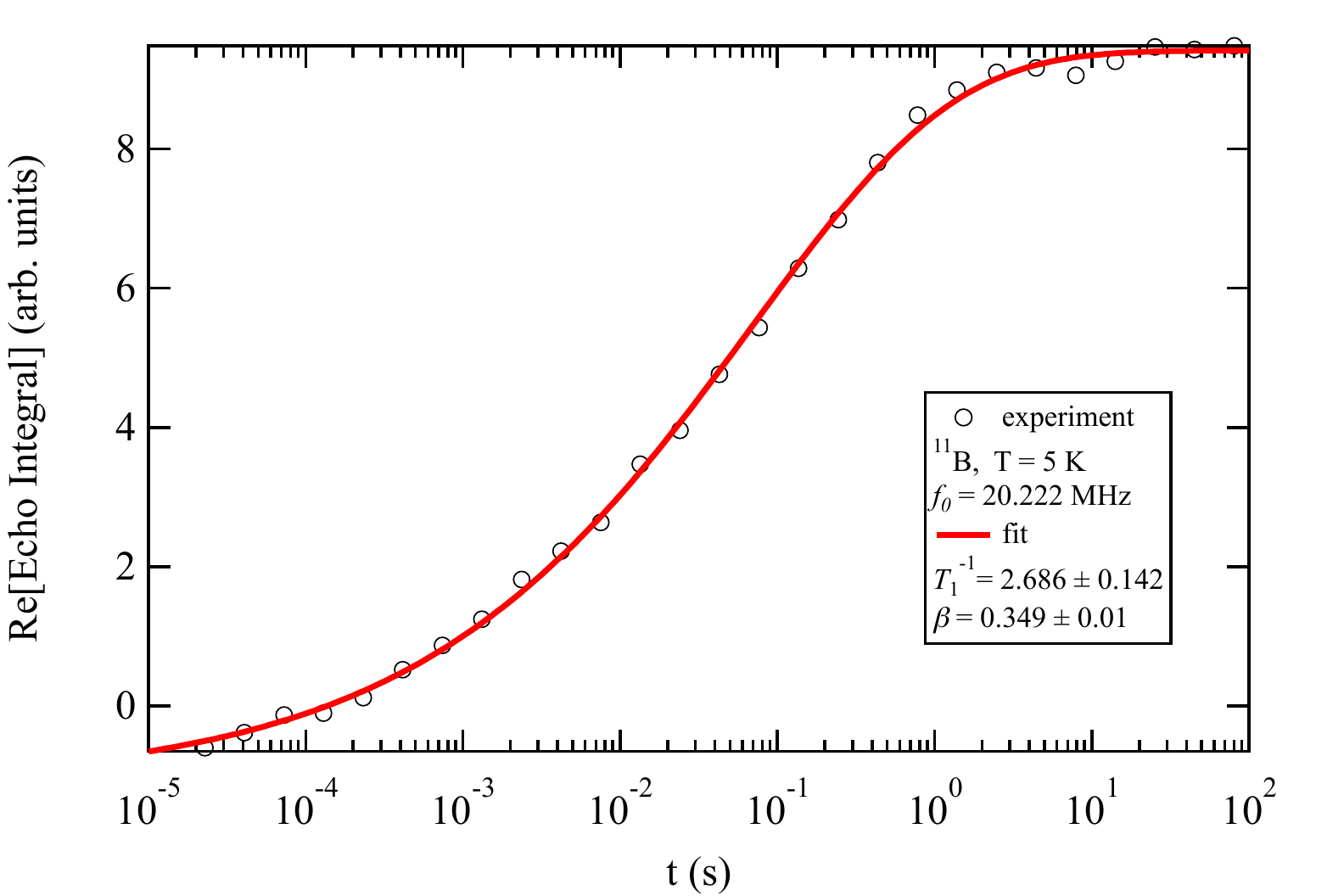}
	\caption{\label{fig:B11_5K_recovery}Nuclear magnetization saturation recovery curve for ${}^{11}$B superposed central transition at $T=5$ K with stretched exponential fit to Eqn.~\ref{eqn:I_3_over_2_strexp}.}
\end{figure}

The spin-lattice relaxation rate of ${}^{11}$B was measured at the central transition of the powder pattern. The corresponding normal modes relaxation equation for inversion/saturation of the central transition of ${}^{11}$B is given by,
\begin{equation}
\label{eqn:I_3_over_2_strexp} 
\begin{split}
	M_N(t) = M_N(\infty)\big(1 - {\alpha}\Big(& \frac{9}{10}e^{-(6t/T_1)^\beta} \\
	&+ \frac{1}{10}e^{-(t/T_1)^\beta}\Big)\big).
\end{split}
\end{equation}
Because the ${}^{11}$B sites all have roughly the same shift and similar EFG, we find that the central transitions overlap as shown in Fig.~\ref{fig:B11_spectra_layout}(b). This results in a situation where deviations from $\beta=1$ in Eq.~\ref{eqn:I_3_over_2_strexp} could be caused by different relaxation at the three ${}^{11}$B sites and/or anisotropy of the relaxation rate. The fact that the stretching exponent ${}^{11}\beta \sim 0.35$ (fit shown in Fig.~\ref{fig:B11_5K_recovery}) is smaller than ${}^{239}\beta \sim 0.6$ indicates that this spectral overlap likely contributes to deviation from the expected relation function. Single crystal studies where one can measure $T_1$ at a well-separated satellite transition of a single site are required to extract the precise value of $T_1$ for ${}^{11}$B. As the focus of this work is the ${}^{239}$Pu, we chose to study a powder which enables extraction of both ${}^{239}K_c$ and ${}^{239}K_a$ from a single powder pattern at each temperature.

In spite of this overlap, we find that the relaxation rate distribution of the ${}^{239}$Pu is peaked roughly an order of magnitude faster than that of the ${}^{11}$B relaxation rate distribution, indicative of the difference in strength of the hyperfine fields at the nuclei. The magnitude of the relaxation rates reflect that the Pu 5$f$-electrons are in a non-magnetic configuration. If the 5$f$-electrons were magnetic, then fluctuations of the transferred hyperfine fields at the ${}^{11}$B site would drive fast much faster relaxation of the ${}^{11}$B, and fluctuations of the on-site hyperfine field would drive extremely fast relaxation at the ${}^{239}$Pu site such that the NMR signal would relax too quickly to observe.

\bibliography{PuB4_NMR_SM}